\documentclass[12pt]{article}
\usepackage{epsfig}
\usepackage{amsfonts}
\usepackage{amsopn}
\usepackage{color}
\usepackage{epic}

\newcommand{\eq}{\begin{equation}}
\newcommand{\eqx}{\end{equation}}
\newcommand{\eqn}{\begin{eqnarray}}
\newcommand{\eqnx}{\end{eqnarray}}
\newcommand{\f}[2]{\frac{#1}{#2}}

\renewcommand{\th}{\theta}
\newcommand{\sg}{\sigma}
\newcommand{\dl}{\delta}
\newcommand{\chit}{\tilde{\chi}}
\newcommand{\xtl}{\tilde{x}}
\newcommand{\nn}{{\cal N}}

\title{From nesting to dressing} 

\author{Romuald A. Janik\thanks{e-mail: {\tt ufrjanik@if.uj.edu.pl}}
\  and Tomasz {\L}ukowski\thanks{e-mail: {\tt tomaszlukowski@gmail.com}} \\ \\
Institute of Physics\\
Jagellonian University,\\
ul. Reymonta 4, \\
30-059 Krak{\'o}w\\
Poland
}

\begin{document}

\maketitle

\begin{abstract}
In integrable field theories the S-matrix is usually a product of a relatively
simple matrix and a complicated scalar factor.
We make an observation that in many relativistic integrable field
theories the scalar factor can be expressed as a convolution of simple
kernels appearing in the nested levels of the nested Bethe ansatz. 
We formulate a proposal, up to some discrete ambiguities, how to reconstruct
the scalar factor from the nested Bethe equations and check it for
several relativistic integrable field theories. We then apply
this proposal to the AdS asymptotic Bethe ansatz and recover the
dressing factor in the integral representation of Dorey, Hofman and
Maldacena.
\end{abstract}

\section{Introduction}

The AdS/CFT correspondence \cite{adscft}, which links $\nn=4$ Super
Yang-Mills theory and $AdS_5 \times S^5$ superstring theory provides a
fascinating and powerful method for studying the behaviour of gauge
theory at strong coupling. Yet that same feature makes it very
difficult to test in the whole range of coupling  constants. If we
want to make contact with perturbative gauge theory we are forced to
investigate the deeply quantum regime of string theory. On the other
hand if we start from gauge theoretical constructions, passage to
strong coupling is very difficult.

The discovery of integrability
\cite{Minahan:2002ve,Beisert:2003tq,Beisert:2003yb,kor1,Bena:2003wd,Dolan:2003uh}
on both sides of the 
correspondence gave new tools for making the strong-weak coupling
interpolation. On the gauge theory side integrable spin chains have been
investigated for general operators at 1-loop \cite{BS0}. The
integrable structures on both sides of the correspondence were
analyzed from the point of view of finite-gap solutions \cite{KMMZ}. 
Then the spin chain formulation was extended to an
all-loop asymptotic Bethe ansatz \cite{BDS,S,BS}. Yet it turned that
in order to describe correctly strong coupling string physics one has
to introduce a scalar function -- the so-called `dressing factor'
\cite{AFS,BT,HL} which was unconstrained by integrability. 

In parallel, the S-matrix (which could be interpreted both as an
ingredient of the spin chain language or as a worldsheet S-matrix for
superstring excitations) was found to be determined by symmetry again
up to the scalar factor \cite{B}. Despite the lack of relativistic
invariance, constraints from crossing symmetry were derived
\cite{CROSS}, and a dressing factor(s) solving these constraints was
found \cite{BHL}. Later in \cite{BES} a specific member of the class
of solutions was choosen.

One of the intriguing features of the form of the dressing factor was
its extreme complexity. Even though integral representations have been
found \cite{KLip,Didina,DHM}, the complicated formulas prompted
some conjectures that the dressing factor arises dynamically from some
hidden levels of a more fundamental Bethe ansatz \cite{ND}. This idea
was earlier partly realised in some models \cite{MP,Kaz,KG}, and other
works \cite{Sakai} suggested a construction which works for large length. 
However no completely satisfactory construction along these lines has
been found so far. 

The aim of this paper is to show that the complexity of the AdS
dressing factor is only apparent and that it shares a lot of its
structure with quite simple ordinary relativistic integrable field
theories. We make an observation that the nested structure of Bethe
ansatz is intertwined with the form of the scalar factor and we found
a `phenomenological' procedure (up to a set of discrete ambiguities)
of constructing the dressing factor out of simple ingredients entering
the nested levels. The outcome is a generic structure of the scalar
factor as (multiple) convolution of simple kernels appearing in the
nested Bethe equations very much reminiscent of the Dorey, Hofman,
Maldacena integral representation of the AdS dressing factor.

The plan of this paper is as follows. In section 2 we review the basic
facts about the BHL/BES dressing factor and quote the integral
expression of \cite{DHM}. In section 3 we describe thermodynamic Bethe
ansatz (TBA) for relativistic integrable field theories and analyze
the case of $(RSOS)_3$ model which is the basis for our
observation. Then we summarize our procedure and check that it works
for the $O(4)$ model and more generally for all $O(2n)$ models. Then
in section 4 we move to the case of $AdS_5 \times S^5$ and show how
one can recover the integral formula of Dorey, Hofman and Maldacena
from our construction which has as its input just the asymptotic Bethe
ansatz. We close the paper with a conclusion and two appendices.

\section{The BHL/BES dressing factor}

The dressing factor of the $AdS_5 \times S^5$ superstring worldsheet
theory is currently believed to be given by the BHL/BES
expression \cite{BHL,BES}. This form satisfies all the known
constraints both at weak 
and at strong coupling as well as it ensures that the S-matrix has
crossing symmetry \cite{CROSS}. Its general form is given by the factorization
\eq
\sg^2(x_q,x_p)=\f{R^2(x_q^-,x_p^-) R^2(x_q^+,x_p^+)}{ R^2(x_q^-,x_p^+)
  R^2(x_q^+,x_p^-)} 
\eqx
where each of the factors is given by
\eq
R(x,y)=e^{i\chi(x,y)}
\eqx
where $\chi(x,y)=\tilde{\chi}(x,y)-\tilde{\chi}(y,x)$ is antisymmetric
and defined through the series expansion \cite{BHL,BES}
\eq
\label{e.defchin}
\chit(x,y) =\sum_{r=2}^\infty \sum_{s=r+1}^\infty
\f{-c_{r,s}(g)}{(r-1) (s-1)} \f{1}{x^{r-1} y^{s-1}}
\eqx
where $c_{r,s}(g)$  have the strong coupling expansion
\eq
c_{r,s}(g)=\sum_{n=0}^\infty c^{(n)}_{r,s} \f{1}{g^{n-1} }
\eqx  
with the coefficients $c^{(n)}_{r,s}$ given by 
\eq
c^{(0)}_{r,s}=\dl_{r+1,s}
\eqx
for the leading AFS part \cite{AFS},
\eq
c^{(1)}_{r,s}=\f{(-1)^{r+s}-1}{\pi} \cdot \f{(r-1)(s-1)}{(r+s-2)(s-r)}
\eqx
for the 1-loop HL part\footnote{Based in particular on the 1-loop
  computations in \cite{BT}.} \cite{HL} and the BHL/BES choice \cite{BHL,BES}
\eq
c^{(n)}_{r,s}=\f{(1-(-1)^{r+s}) \zeta(n)}{2 (-2\pi)^n \Gamma(n-1)}
(r-1)(s-1) \f{\Gamma(\f{1}{2} (s+r+n-3)) \Gamma(\f{1}{2} (s-r+n-1))}{
\Gamma(\f{1}{2} (s+r-n+1)) \Gamma(\f{1}{2} (s-r-n+3))}
\eqx
for $n \geq 2$

The above strong coupling expansion is only asymptotic. In \cite{BES}
there appeared a corresponding expansion at weak coupling
\eq
c_{r,s}(g)=\sum_{n=0}^\infty \tilde{c}^{(-n)}_{r,s} g^{1+n}
\eqx
which could be evaluated to give
\eq
\label{e.crsweak}
c_{r,s}(g) =2\cos \left( \f{1}{2} \pi (s-r-1) \right) \cdot
(r-1)(s-1) \int_0^\infty dt \f{J_{r-1}(2gt) J_{s-1}(2gt)}{t(e^t-1)} 
\eqx
which agrees
with the asymptotic strong coupling formulas quoted above
\cite{FIORAVANTI}. Later a convenient integral expression based on
(\ref{e.crsweak}) was derived
by Dorey, Hofman and Maldacena \cite{DHM}
\eqn
\label{e.dhm}
\chit_{DHM}(x,y)&=&-i\int_{C_1} \f{dz_1}{2\pi} \int_{C_2} \f{dz_2}{2\pi}
\f{1}{x-z_1} \f{1}{y-z_2}\cdot \nonumber\\
&& \cdot \log\Gamma\left(
1+ig \left(z_1+\f{1}{z_1}-z_2-\f{1}{z_2} \right)\right) 
\eqnx 
which resums all terms in the weak coupling expansion\footnote{Other
  forms of integral expressions for the dressing kernel have been
  derived earlier in \cite{KLip,Didina}, but the one from \cite{DHM}
  will be most convenient for our purposes.}. In the above expression
the integration contours $C_i$ 
are unit circles. Even though the above expression is much simpler
than the rather complicated expressions for the dressing factor
coefficients it is still rather formidable. One of the main points of this
paper is to show that the above structure is in fact quite natural. In
the next section we will return to conventional relativistic
integrable field theories and show that the scalar factors for a wide
range of theories with {\em nondiagonal} scattering have a very
similar form.   

\section{Integrable relativistic theories -- TBA}

The point of departure of our observation was the rather surprising
simplicity of thermodynamic Bethe equations for the simplest
relativistic integrable theory with nondiagonal scattering: the $(RSOS)_3$
model. Before we describe it in detail let us briefly review the
thermodynamic Bethe ansatz (TBA) in the case of diagonal scattering
and a single species of particles \cite{zamTBA}. 

The aim of the thermodynamic Bethe ansatz in its original form is to
compute the ground state energy $E_0(L)$ for an integrable quantum field
defined on a cylinder of circumference $L$. The
ground state energy can be reconstructed from the euclidean partition
function
\eq
E_0=-\lim_{R\to \infty} \f{1}{R} \log  Z(R,L)
\eqx
The direct computation of this partition function in an interacting,
even integrable, theory is very complicated due to virtual particles
going around the cylinder and interacting with each other. The
Thermodynamic Bethe Ansatz amounts
to computing the same partition function treating $R$ as space and $L$
as a compactified time, i.e. inverse temperature. Now since the space
coordinate is decompactified the complications due to virtual
particles go away and one can use ordinary Bethe ansatz quantization:
\eq
e^{imR \sinh \th_i} \prod_{j \neq i} S(\th_i-\th_j)=1
\eqx
which, after taking logarithms, becomes
\eq
\label{e.bethelog}
m R\sinh\th_i +\sum_{j\neq i} \dl(\th_i-\th_j)=2\pi n_i
\eqx
where $n_i$ is an integer and $\dl=-i\log S$. If we look at a specific
solution to the system (\ref{e.bethelog}), we find that it corresponds
to a set of, non-necessarily consecutive, integers. We will call those
roots `occupied roots'.
Then because $R \to \infty$, we are led to consider continous
distributions of roots. Taking the derivative w.r.t. $\th_i$ we thus
obtain
\eq
\label{e.tbadiag}
\rho(\th)+\rho^h(\th)=\frac{1}{2\pi}m R \cosh \th +\int \phi(\th-\th')
\rho(\th') d\th'
\eqx
where $\rho(\th)$ is the density of occupied roots, $\rho^h(\th)$ is
the density of {\em unoccupied} roots (`holes') while $\phi(\th)$ is
the S-matrix kernel
\eq
\phi=\frac{1}{2\pi i}\partial_\th \log S
\eqx
At this stage, equation (\ref{e.tbadiag}) is a relation between two unknown
quantities -- the densities of particles and holes.
In the further steps of the TBA procedure one minimizes the free
energy obtaining a second equation which links $\rho(\th)$ to
$\rho^h(\th)$ thus forming, together with (\ref{e.tbadiag}), a closed system of
equations which determines $\rho$ and $\rho^h$ and consequently the
partition function and the ground state energy $E_0(L)$.

For our applications, it is already a feature of the
equation  (\ref{e.tbadiag}) written with $\rho$ and $\rho^h$ taken to
be independent which will be at the basis of our proposal. Nothing
will depend on the further steps of TBA so we will not describe them
here in more detail.

\subsection{The $(RSOS)_3$ model}

In this section we would like to consider, following \cite{zamRSOS},
the analogues of the  equation (\ref{e.tbadiag}) for one of the
simplest theories 
with nondiagonal scattering -- the $(RSOS)_3$ model. Since the
S-matrix is now nondiagonal, 
instead of the ordinary Bethe ansatz one has to deal with additional
levels of nesting and introduce additional species of roots which do
not carry any momentum or energy but which encode the diagonalization
of the transfer matrix.

The S-matrix of the RSOS model has the following structure:
\eq
\label{e.srsos}
\hat{S}=\f{1}{\scriptstyle \cosh^{\f{1}{2}}\th } \cdot 
\underbrace{e^{\f{i}{4} \int_0^\infty
    \f{dt}{t} \f{\sin \f{t \th}{\pi}}{\cosh^2 \f{t}{2}}}}_{\sg(\th)}
\cdot \left(\parbox{2.5cm}{Trigonometric functions of $\th$}\hspace{0cm}\right) 
\eqx
We see here that the scalar factor $\sg(\th)$ necessary for imposing crossing
invariance is much more complicated than the trigonemetric functions
appearing in the matrix structure. This is a generic situation in
theories with nondiagonal scattering.

When applying Bethe ansatz quantization to the S-matrix (\ref{e.srsos})
we have one additional flavour of roots:
$x_i$ which come in two varieties
\eq
x_i^\pm =y_i \pm i\f{\pi}{2}
\eqx
The $y_i$'s satisfy the equation
\eq
\prod_{k=1}^N \f{\sinh \left(\f{y_i-\th_k}{2}+i\f{\pi}{4}\right)}{
\sinh \left(\f{y_i-\th_k}{2}+i\f{\pi}{4}\right)}= \pm 1
\eqx
where the corresponding imaginary parts of $x_i$'s may be either
$+i\pi/2$ or $-i\pi/2$.
The `physical' momentum carrying roots satisfy the equation
\eq
\label{e.rsosmid}
e^{im R \sinh \th_k} \prod_l \sg(\th_k-\th_l) \prod_i \sinh
\f{\th_k-x_i}{2} \cdot (\ldots)=1
\eqx
where $(\ldots)$ are terms which do not contribute in the thermodynamic
limit. Note the appearance of the complicated 
`dressing phase' $\sg(\th)$. In \cite{zamRSOS}, Zamolodchikov derived
the continous equations of this model in the course of applying the
TBA procedure. The first equation is
\eq
\label{e.nestrsos}
\rho^+(y)+\rho^-(y)=\frac{1}{2\pi}\int\f{1}{\cosh(y-\th)} \rho(\th) d\th
\equiv K *\rho
\eqx
which involves the rather simple kernel $K=\f{1}{2\pi}
\f{1}{\cosh(y-\th)}$ coming from 
the nested level equation. The second equation quoted by Zamolodchikov
reads
\eqn
\label{e.zam}
\rho(\th)+\rho^h(\th) &=& \frac{1}{2\pi}m R \cosh \th 
+\frac{1}{2\pi}\int\f{1}{\cosh(\th-y)} 
\rho^+(y) dy \nonumber\\
&\equiv& \frac{1}{2\pi}m R \cosh \th+ K *\rho^+
\eqnx
Let us emphasize that this equation looks extremely surprising. It
came from taking the continuum limit of (\ref{e.rsosmid}) and so
should incorporate somehow the dressing kernel ($\phi_\sg \equiv
\frac{1}{2\pi i}\partial_\th \log \sg$). Yet this equation only
involves the much 
simpler `nesting kernel' $K$. 

In order to see in detail how this came about let us write
(the imaginary part\footnote{The real part is also satisfied -- see
  \cite{zamRSOS}.} of) the continuum version of (\ref{e.rsosmid}):
\eq
\rho+\rho^h= \frac{1}{2\pi}m R \cosh \th +\phi_\sg * \rho+\f{1}{4\pi}
\int\f{\rho^+(y)-\rho^-(y)}{\cosh(\th-y)} dy
\eqx  
Now we may express $\rho^-(y)$ from (\ref{e.nestrsos}) and plug it
into the above equation. We obtain 
\eq
\rho+\rho^h=\frac{1}{2\pi} m R \cosh \th +\phi_\sg * \rho-\f{1}{2} K*K*\rho
+K * \rho^+
\eqx
This is indeed equal to (\ref{e.zam}) once we note that the specific
dressing phase of the RSOS model satisfies
\eq
\phi_\sg =\f{1}{2} K*K
\eqx 
This surprising expression is the main motivation of our work. It
gives an expression for the dressing kernel, which  is a complicated
looking transcendental function, as a convolution of much simpler
kernels which come from nested levels of Bethe equations. The
appearance of such a formula suggests that the complicated form of the
dressing phases (S-matrix scalar factors)  may just be a byproduct of
some simple structure extracted from the `simple' matrix part of the
S-matrix.

\subsection{The proposal}

The surprising cancelation observed in the case of TBA equations for
the $(RSOS)_3$ model
prompted us to conjecture that such a phenomenon may be 
more general. If we abstract the basic steps leading to the
cancelation the following procedure suggests itself. 

First consider the momentum carrying equation in the Bethe ansatz and
write it in the continuum limit as for TBA. In the case of nondiagonal
scattering this equation will involve auxiliary Bethe roots involved
in the diagonalization of the transfer matrix:
\eq
\ldots +\sum_i K_i * \rho_i
\eqx
where $\rho_i$ are the densities of the auxiliary roots. Then we
use the nested levels of the Bethe equations to express the $\rho_i$'s
in terms of the density of the momentum carrying roots $\rho$:
\eq
\rho_i=\tilde{K}_i * \rho+\ldots
\eqx 
where $\ldots$ may stand for e.g. densities of holes etc. Then the
observation made for the $(RSOS)_3$  model suggests that the dressing
kernel has an expression of the form
\eq
\label{e.guess}
K_\sg =-\sum_i K_i * \tilde{K}_i
\eqx
The origin of this `phenomenological' observation is unclear to us. We
suspect that it must be a feature of some internal consistency between
the structural properties of the nested Bethe ansatz and crossing
which fixes the overall scalar factor of the S-matrix which we call
here generically `the dressing factor'. Note that there is some
ambiguity in the above construction. The nested equations could be
inverted which would lead to various sign changes in the kernels
(which arise through taking the logarithms of the equations -- this is
especially evident in the case of $O(2n)$ models to be discussed below). We do
not know what fixes the correct signs -- what is surprising for us is
that such a choice {\em exists} at all. In order to check whether this
is not just a coincidence, we will 
analyse in the next section the $O(4)$ model from the same
perspective and then we will verify that the same procedure works also
for all $O(2n)$ models.

Then in section 4, we will proceed to use the same mechanism for the
$AdS_5 \times S^5$ superstring/$\nn=4$ SYM Bethe ansatz.

\subsection{$O(4)$ model}

The $O(4)$ model is described by the  S-matrix
\eq
\hat{S}^{cd}_{ab}(\th)=\sg^2(\th) \f{\th}{\th-i} \left[ \dl^c_a
\dl^d_b- \f{i}{\th} \dl_a^d \dl_b^c -\f{i}{i-\th} \dl_{ab} \dl^{cd} \right]
\eqx
where the scalar factor has the integral representation
\eq
\label{e.ofour}
\sg^2(\th)=e^{2i \int_0^\infty \f{dk}{k} \sin k\th \cdot
  \f{2e^{-k}}{1+e^{-k}} }
\eqx
alternatively it can be expressed as a product of Gamma functions \cite{zamON}.

Let us quote the Bethe equations of the $O(4)$ model. Apart from
the physical $\th_k$'s we have two sets of auxillary roots: $u_i$'s
and $v_j$'s. There are three Bethe equations:
\eqn
1 &=& \prod_k \f{u_j-\th_k-\f{i}{2}}{u_j-\th_k+\f{i}{2}} \prod_{i\neq j}
\f{u_j-u_i+i}{u_j-u_i-i} \\
e^{-im\sinh \pi \th_k} &=& \prod_{j \neq k} \sg^2(\th_k-\th_j) \prod_l
\f{\th_k-u_l+\f{i}{2}}{\th_k-u_l-\f{i}{2}}  
\prod_l \f{\th_k-v_l+\f{i}{2}}{\th_k-v_l-\f{i}{2}}  \\
1 &=&  \prod_k \f{v_j-\th_k-\f{i}{2}}{v_j-\th_k+\f{i}{2}} \prod_{i\neq j}
\f{v_j-v_i+i}{v_j-v_i-i}
\eqnx
Let us now define the continuum limit of these equations which in
principle would be the starting point of TBA\footnote{Note however
  that the TBA equations for $O(4)$ models are more complicated \cite{BH0,BH}.}.
It is convenient to use the notation

\eqn\label{phi.ofour}
\phi_{a}(x)&\equiv &\frac{1}{2\pi i} \partial_{x}\log(\frac{x+i a}{x-i
  a})=-\frac{1}{\pi}\frac{a}{x^2+a^2}\\ 
\label{phisg.ofour}
\phi_{\sg}(\th)&\equiv &\frac{1}{2\pi i}\partial_{\th}\log(\sg(\th))
\eqnx

We obtain
\eqn
\rho_u+\rho_u^h &=& \phi_{-\f{1}{2}} * \rho +\phi_1 * \rho_u \\
\rho+\rho^h &=& \frac{1}{2\pi}mR \cosh \th +2\phi_\sg *\rho +
\phi_{\f{1}{2}}*\rho_u 
+ \phi_{\f{1}{2}}*\rho_v \\
\rho_v+\rho_v^h &=& \phi_{-\f{1}{2}} * \rho +\phi_1 * \rho_v
\eqnx
Now let us perform similar manipulations as in the $(RSOS)_3$ model. First
let us express the auxillary densities $\rho_u$ and $\rho_v$ in terms
of $\rho$. We obtain
\eq
\rho_{u,v}=(1-\phi_1)^{-1} * \phi_{-\f{1}{2}}* \rho +\ldots
\eqx
where the dotted terms $\ldots$ contain only hole densities. Then
according to our 
proposal we would expect the dressing kernel to be given by the
formula 
\eq
\phi_\sg=-\phi_{\f{1}{2}} * (1-\phi_1)^{-1} * \phi_{-\f{1}{2}}
\eqx
This convolution can be easily evaluated using Fourier transforms and
checked that the result indeed exactly coincides with
(\ref{e.ofour})! So we see that again the dressing kernel arises as
convolution of simpler nested kernels which come from the nested
levels of the Bethe equations. In fact the above decomposition is
structurally reminiscent of the DHM formula (\ref{e.dhm}) -- it is
also given by a double convolution with the extreme factors being
quite simple.

Since we do not have a real understanding why the above procedure
seems to work, we decided to check it also in the more general case of
$O(2n)$ models - as the $O(4)$ model is really quite special being
essentially a product of two $SU(2)$ S-matrices.

\subsection{$O(2n)$ models}

For $n\geq 2$ the $O(2n)$ $\sigma$-model is described by the S-matrix
which was proposed a long time ago by Zamolodchikov and Zamolodchikov
\cite{zamON} and takes the form 
	\eq\label{smatrix}
	\hat{S}^{cd}_{ab}(\theta)=\sg^2(\th)\frac{\theta}{\theta-i}\Big[
          \delta^{c}_{a}\delta^{d}_{b}-\frac{i}{\theta}\delta^{d}_{a}
          \delta^{c}_{b}-\frac{i}{i(n-1)-\theta} 
          \delta_{ab}\delta^{cd}\Big]  
	\eqx
where $\sg(\th)$ is the dressing factor.
	
For given $n$ we have a tower of Bethe equations, one for
each type of Bethe root, which consists of the main `momentum
carrying' Bethe equation  
	\eq\label{bethe1}
	e^{-i L
          \sinh(\frac{\pi}{n-1}\theta_{\alpha})}=\prod_{\beta\neq
          \alpha}^{K_{n+1}}S_{0}(\theta_{\alpha}-\theta_{\beta})\prod_{j=1}^{K_{n}}
        \frac{\theta_{\alpha}-u_{j}^{(n)}+\frac{i}{2}}{\theta_{\alpha}-
          u_{j}^{(n)}-\frac{i}{2}}           
	\eqx	
and $n$ `nested' equations (one for each auxillary root). This set of
`nested' equations can be expressed in closed form as
	\eq\label{bethe2}
	-1=\prod_{l=1}^{n+1}\prod_{j=1}^{K_{l}}\frac{u_{i}^{(k)}-u_{j}^{(l)}+
          \frac{i}{2}(\alpha_{k}|\alpha_{l})}{u_{i}^{(k)}-u_{j}^{(l)}-
          \frac{i}{2}(\alpha_{k}|\alpha_{l})}
	\eqx

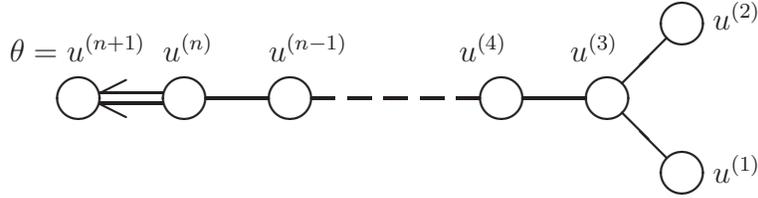
\begin{figure}[t]
\label{dynkin}
%
%
\begin{center}
\unitlength=2.0pt
\begin{picture}(160,50)
\thicklines
\put(24,19){\line(2,-1){5}}
\put(24,21){\line(2,1){5}}
\put(24,19){\line(1,0){12}}
\put(24,21){\line(1,0){12}}
\put(44,20){\line(1,0){12}}
\dashline[50]{4.8}(64,20)(96,20)
\put(104,20){\line(1,0){12}}
\put(122.82,22.82){\line(1,1){8.48}}
\put(122.82,17.18){\line(1,-1){8.48}}
\put(20,20){\circle{8}}
\put(40,20){\circle{8}}
\put(60,20){\circle{8}}
\put(100,20){\circle{8}}
\put(120,20){\circle{8}}
\put(134.14,5.86){\circle{8}}
\put(134.14,34.14){\circle{8}}
\put(7,27){$\theta=u^{(n+1)}$}
\put(36,27){$u^{(n)}$}
\put(56,27){$u^{(n-1)}$}
\put(92,27){$u^{(4)}$}
\put(113,27){$u^{(3)}$}
\put(140,33){$u^{(2)}$}
\put(140,4){$u^{(1)}$}
\end{picture}
\end{center}
\caption{Dynkin diagram for
the $O(2n)$ model}
\end{figure}

Here $\{ \alpha_{k}\}$ for $k=1,\ldots,n$ denote the simple roots of
the $\mathfrak{so}(2n)$ Lie algebra and $(\alpha_{k}|\alpha_{l})$
expresses inner product defined in the root space as 
	\eq
	(\alpha_{k}|\alpha_{l})=\left\{ 
	\begin{tabular}{ll}2& \mbox{for }l=k, (k=1,\ldots,n)\\
	-1&\mbox{for } l=k+1, (k=2,\ldots,n-1)\\
	-1&\mbox{for } k=1, l=3.
	\end{tabular}\right.
	\eqx
Additionally, we introduced an extra root $\alpha_{n+1}$ giving
$(\alpha_{n+1}|\alpha_{k})=-\delta_{nk}$, so that the interaction
between $\theta_{\alpha}$'s and $u_{j}^{(k)}$'s are included in the
common notation (\ref{bethe2}).  The non-zero elements can be encoded
in the Dynkin diagram (Fig.\ref{dynkin}). 
	
Starting from the Bethe ansatz equations (\ref{bethe1}) and
(\ref{bethe2}) we would like to take the continuum limit and transform
them from the discrete to an integral form. From the form of the
Dynkin diagram we see that generic nodes have only two neighbours. For
these Bethe roots we obtain 
	\eq\label{eq.o2n}
	\rho_{k}+\rho^{h}_{k}=\phi_{-\frac{1}{2}}\ast \rho_{k+1}+
        \phi_{-\frac{1}{2}}\ast \rho_{k-1}+\phi_{1}\ast \rho_{k} 
	\eqx
where $\rho_{i}$ denotes the density of $i$-th Bethe roots and
$\rho^{h}_{i}$ the density of holes and $\phi_{a}$ is defined by
(\ref{phi.ofour}). 

Apart from these nodes there are always four nodes which have
different number of neighbours and thus have to be treated separately.
After we take the continuum limit the Bethe equations associated with
these nodes take the form 
	\eqn\label{abethe3}
	\rho_{1}+\rho^{h}_{1}&=&\phi_{-\frac{1}{2}}*\rho_{3}+\phi_{1}\ast
        \rho_{1}\\ \label{abethe4} 
	\rho_{2}+\rho^{h}_{2}&=&\phi_{-\frac{1}{2}}*\rho_{3}+\phi_{1}\ast
        \rho_{2}\\ \label{abethe5} 
	\rho_{3}+\rho^{h}_{3}&=&\phi_{-\frac{1}{2}}*\rho_{1}+
        \phi_{-\frac{1}{2}}*\rho_{2}+\phi_{-\frac{1}{2}}*\rho_{4}+\phi_{1}\ast
        \rho_{3} 
	\eqnx
for the auxillary roots and
\eq
\rho_{n+1}+\rho^{h}_{n+1}=\frac{1}{2(n-1)}\cosh(\frac{\pi}{n-1} \theta_{\alpha})+
2\phi_{\sigma} \ast \rho_{n+1}+\phi_{\frac{1}{2}}\ast 	\rho_{n} 
\eqx
for the main, `momentum carrying' roots. Again we see that this is the
only equation which involves the `dressing factor'.

From the above equation we see, that according to our proposal, we
have to express $\rho_n$ in terms of $\rho_{n+1}$ using
(\ref{eq.o2n})-(\ref{abethe5}) to get $\rho_n=\Phi \ast \rho_{n+1}$
and then the dressing factor should be given by
\eq
\label{e.dro2n}
\phi_\sg =-\f{1}{2}\, \phi_{\frac{1}{2}}\ast \Phi
\eqx

The simplest way to solve the set of equations
(\ref{eq.o2n})--(\ref{abethe5}) is to take the Fourier transform of
it. Doing it we get the set of $n+1$ algebraic equations which is
solved in Appendix B.  As a result we get
	\eq\label{b}
	\hat{\phi}_{\sigma}=\f{1}{2}\hat{\phi}_{-1}
        \frac{1+(\hat{\phi}_{-1})^{n-2}}{
          1+ (\hat{\phi}_{-1})^{n-1}}
	\eqx
and using (\ref{phisg.ofour}) we obtain the formula for the dressing factor
	\eq
	\sg^{2}(x)=\exp\left(2i\int_{0}^{+\infty}\frac{\sin(\frac{\omega\pi}{n-1}
          x)}{\omega}  \frac{e^{-\frac{\omega \pi}{n-1}}+e^{-\omega
            \pi}}{1+ e^{-\omega\pi}}d\omega\right)
	\eqx
what agrees with the result from \cite{BH} taking into account the
different conventions\footnote{Rapidities have to be rescaled
  $\theta\to \frac{\theta (n-1)}{\pi}$, and $n$ is interchanged with $2n$.}.

\subsubsection*{Ambiguities in the choice of signs}

As a note of caution let us point out an ambiguity appearing in our
current proposal. Let us consider e.g. the Bethe equation for the
roots $u^{(1)}_i$ and {\em invert} it to get
\eq
	-1=\prod_{l=1}^{n+1}\prod_{j=1}^{K_{l}}\frac{u_{i}^{(1)}-u_{j}^{(l)}-
          \frac{i}{2}(\alpha_{1}|\alpha_{l})}{u_{i}^{(1)}-u_{j}^{(l)}+
          \frac{i}{2}(\alpha_{1}|\alpha_{l})}
\eqx
This is of course completely equivalent to the previous form. Let us
now proceed to take the thermodynamical limit. The outcome will be 
\eqn
\rho_{1}+\rho^{h}_{1}&=&\phi_{\frac{1}{2}}*\rho_{3}+\phi_{-1}\ast
\rho_{1} \nonumber\\
&=& -\phi_{-\frac{1}{2}}*\rho_{3}-\phi_{1}\ast
        \rho_{1}
\eqnx
Proceeding as before, we will obtain a different formula for the
dressing factor. We do not know how to pick the correct choice on
general grounds. This
means that our proposal has to supplanted by, a yet to be discovered,
criterion. The nontrivial feature coming from all these computations
is that nevertheless a choice exists which reproduces the correct
scalar factor.

\section{Dressing factor from the $AdS_5\times S^5$ Bethe ansatz}

Let us now  apply our proposal to the main case of interest,
namely the $AdS$ S-matrix. As noted before, there is a (finite!) set of choices
of the way one writes the equations which may give different
results. In the relativistic cases of $(RSOS)_3$ or $O(2n)$ models there was
a choice which reproduced the correct scalar factor. 
The same
situation -- a possibility of different choices -- happens in the case
of the asymptotic Bethe ansatz. Since we 
do not know a general proof of our proposal we do not know {\em
  a-priori} how to justify a particular choice. What is surprising is that a
choice exists which leads to the DHM integral formula (\ref{e.dhm}).

Let us start with the Beisert-Staudacher asymptotic Bethe ansatz
written in the form of 7 equations with the grading corresponding to
the $sl(2)$ sector. Let us first look at the momentum carrying node
equation, which again is the only equation involving the dressing
factor. The momentum carrying roots are $x_4^\pm$, which are
intertwined in this equation with roots of type $x_1$, $x_3$, $x_5$
and $x_7$ through the terms:
\eq
\prod_k \f{x_4^- -\f{1}{x_{1,k}}}{x_4^+-\f{1}{x_{1,k}}} \prod_l
\f{x_4^- -x_{3,l}}{x_4^+-x_{3,l}} \cdot \left( x_1 \leftrightarrow
x_7,  x_3 \leftrightarrow x_5 \right)
\eqx
Hence the associated kernels appearing in this equations are
\eq
K_{41} * \rho_1 +K_{43} * \rho_3 + \left( 1 \leftrightarrow 7, 3
\leftrightarrow 5 \right) 
\eqx
Since we have direct symmetry between the roots coming from the two
sides of the Dynkin diagram from now on we will just consider the
contribution of the roots $x_{1,2,3}$.

Now in order to use our proposal (\ref{e.guess}) we have to solve the
first three equations for the densities $\rho_1$ and $\rho_3$ and
express them in terms of $\rho_4$ and the hole densities -- of course
only the part proportional to $\rho_4$ will be relevant for us.

Let us write these three equation in the following form:
\eqn
1 &=& \prod_j \f{u_1-u_{2,j}-\f{i}{2g}}{u_1-u_{2,j}+\f{i}{2g}} \prod_k
\f{\f{1}{x_1}-x^-_{4,k}}{\f{1}{x_1}-x^+_{4,k}} \\
1 &=& \prod_j \f{u_2-u_{2,j}-\f{i}{g}}{u_2-u_{2,j}+\f{i}{g}} \prod_l
\f{u_2-u_{3,l}+ 
  \f{i}{2g}}{u_2-u_{3,l}-\f{i}{2g}} \prod_m  \f{u_2-u_{1,m}+
  \f{i}{2g}}{u_2-u_{1,m}-\f{i}{2g}} \\
1 &=& \prod_j \f{u_3-u_{2,j}+\f{i}{2g}}{u_3-u_{2,j}-\f{i}{2g}} \prod_k
\f{x_3-x^+_{4,k}}{x_3-x^-_{4,k}}
\eqnx
where 
\eq
u_i \equiv x_i+\f{1}{x_i}
\eqx

We will now proceed to rewrite these equations in the continuous
form:
\eqn
\rho_1+\rho_1^h&=&\phi_{-\f{1}{2g}} *\rho_2 +K_{14} * \rho_4 \\
\rho_2+\rho_2^h&=&\phi_{-\f{1}{g}} *\rho_2 +\phi_{\f{1}{2g}} *\rho_3 +
\phi_{\f{1}{2g}} *\rho_1 \\ 
\rho_3+\rho_3^h&=&\phi_{\f{1}{2g}} *\rho_2 -K_{34} * \rho_4
\eqnx
Note that here the convolutions with the $\phi_a$ kernels are defined
w.r.t. the $u_{1,2,3}$ variables. Solving first for $\rho_2$ gives
\eq
\rho_2= (1-\phi_{-\f{1}{g}})^{-1} * \phi_{\f{1}{2g}} * (\rho_1+\rho_3) +\ldots
\eqx
where $(\ldots)$  here and below stands for terms with hole densities which
will not be relevant for our proposal. Plugging the above equation into the
remaining ones we obtain easily
\eqn
\rho_1 &=& -L * (K_{14}-K_{34}) *\rho_4+ K_{14}*\rho_4+\ldots \\
\rho_3 &=& L * (K_{14}-K_{34}) * \rho_4 -K_{34} * \rho_4 +\ldots
\eqnx
where $L$ is the kernel
\eq
\label{e.l}
L= \phi_{\f{1}{2g}} *  (1-\phi_{-\f{1}{g}})^{-1} * \phi_{\f{1}{2g}}
\eqx
Our proposal (\ref{e.guess}) then gives the following expression for
the dressing kernel: 
\eq
\label{e.dressi}
\phi_\sg=-(K_{41} *K_{14}-K_{43}*K_{34}- (K_{41}-K_{43}) * L * (K_{14}-K_{34}))
\eqx
In order to make this formula concrete we have to pick a choice of
contours of integration for the convolutions appearing in
(\ref{e.dressi}). 

\subsection*{Contours}

In the above expression the convolution is in terms of $x$ and not in
terms of $u=x+1/x$, while in the nested levels we can deal
with the kernel only by taking convolution w.r.t. $u$ on the whole
real line $u\in (-\infty,\infty)$. Hence we have to pick a
corresponding contour in the $x$ plane. There are basically two
natural choices: (i) integral from $-\infty$ to $-1$, then an upper or
lower semicircle and then from $1$ to $\infty$, or (ii) integral
from 0 to $-1$, then a semicircle to $1$ and then the interval from
$1$ to 0 along the positive real axis.

In order to have a resulting effective closed contour we will pick the
contour (ii) for $x_1$ and the contour (i) for $x_3$. Then the
convolution with the kernel $K_{41}-K_{43}$ is effectively just a
convolution along {\em the unit circle} since
\eq
\int_{C_1} dx \partial_x \log \f{x_4^--\f{1}{x}}{x_4^+-\f{1}{x}}(\ldots) -
\int_{C_3} dx \partial_x \log \f{x_4^--x}{x_4^+-x}(\ldots) = 
-\oint_{|x|=1} dx  \partial_x \log \f{x_4^--x}{x_4^+-x}(\ldots)
\eqx
Similarly combining the two first terms $K_{41} *K_{14}-K_{43}*K_{34}$
together, we see that the sum  
becomes just an integral over the unit circle. Moreover, since for
physical particles we have $|x_4^\pm|>1$, these integrals {\em vanish}.

\subsection*{Structural properties of the dressing factor}

Using the above choice of contours, we thus get the following expression of
the dressing kernel
\eq
\label{e.dressk}
\phi_{\sg}=\f{1}{4\pi^2}\oint_{|x|=1} dx \oint_{|y|=1} dy  \partial_z
\log \f{x_4^+(z)-x}{x_4^-(z)-x} L(x,y) 
\partial_y \log \f{\xtl^+_4-y}{\xtl^-_4-y}
\eqx
The variable $z$ is some parameter for the momentum carrying roots
which is not essential since in any case we will be integrating later
w.r.t. $z$ in order to get directly the dressing phase.

Let us note now some structural properties of the the above
expression. First because of the logarithm, the terms with $x^+$
enter (\ref{e.dressk}) with opposite sign as the terms with $x^-$. 
This means that
the dressing phase factorizes in exactly the expected form:
\eq
\label{e.fact}
\sg(x_4^\pm,\xtl_4^\pm) =\f{R(x_4^+,\xtl_4^+)R(x_4^-,\xtl_4^-)}{
  R(x_4^-,\xtl_4^+) R(x_4^+,\xtl_4^-)}
\eqx

Another very important property of the dressing factor is its
antisymmetry. A simple property of the kernel $L(x,y)$ is that its
Fourier transform is a function of the modulus $|k|$. Hence the
integral over $k$ splits naturally into two parts. After a redefinition
of variables and an integration by parts (see formula
(\ref{ads.antisym}), antisymmetry follows.

Finally let us note that the kernel $L(x,y)$, which is obtained from
the nested kernels depends on $x$ and $y$ only through the combination
\eq
x+\f{1}{x}-y-\f{1}{y}
\eqx
exactly as in the Dorey-Hofman-Maldacena integral formula.

\subsection*{The dressing factor}

In order to obtain an expression directly for the dressing factor, we
have to integrate w.r.t. the $z$ parameter, and also, in order to get
a symmetric expression, integrate by parts w.r.t. $y$. Then using the
factorization (\ref{e.fact}) we get the expression for
$\chi(x_4^+,\xtl_4^+)$ 
\eq
\label{ads.antisym}
\frac{2\pi }{4\pi^2}\oint_{|x|=1} dx \oint_{|y|=1} dy
\f{1}{x^+_4-x} \int_{-\infty}^\infty  \f{dk}{2\pi ik} L(k) 
e^{-ik \cdot \left( x+\f{1}{x}-y-\f{1}{y} \right)}  \f{1}{\xtl_4^+-y}
\eqx
where $L(k)$ are the Fourier components of (\ref{e.l}) which can be
easily evaluated using the definition (\ref{phi.ofour}) (keeping in
mind all the conventions from Appendix~A) to give 
\eq
L(k)=\f{1}{e^{\f{|k|}{g}}-1}
\eqx
Now using the formula
\eq
\int_0^\infty \f{dk}{k} \f{e^{-ik z}}{e^k-1} =C_1+iz C_2+\log
\Gamma(1+i z)
\eqx
we see that our expression reduces exactly to the
Dorey-Hofman-Maldacena integral representation:
\eq
-i\oint_{|x|=1} \f{dx}{2\pi} \oint_{|y|=1} \f{dy}{2\pi} \f{1}{x^+_4-x}  \f{1}{\xtl_4^+-y} \log \Gamma
\left(1+ig \left(x+\f{1}{x}-y-\f{1}{y}\right)\right)
\eqx

\section{Conclusions}

In this paper we made an observation that for a wide range of
relativistic integrable field theories, the overall scalar factor of
the S-matrix which makes the S-matrix crossing invariant can be
expressed as a convolution of simple kernels appearing in the nested
Bethe ansatz. In this way the complicated structure of this scalar
factor just comes about from convolutions of simple ingredients. This
must be a reflection of some, yet to be discovered, structural
self-consistency of nested Bethe ansatz and crossing symmetry. 

We argue that if we assume such property to be general, we may
reconstruct, up to a discrete set of choices, the extremely
complicated BHL/BES dressing factor just starting from the asymptotic
Bethe ansatz of \cite{BS}. The structure of the dressing phase appearing as a
double convolution of various kernels is a direct counterpart of
similar formulas appearing in completely conventional relativistic
integrable field theories.

Let us comment on the present observation in relation to various
proposals which have been made concerning the understanding of the AdS/CFT
dressing phase. It has been suggested that the immensely complicated
structure of the dressing factor suggests that it arises from some
hidden levels of some more fundamental Bethe ansatz (\cite{ND},
see also \cite{MP,Kaz,KG}). Alternatively,
that it could arise from scattering around some filled new vacuum
state\footnote{Although such interpretation is problematic for short
  operators/lengths.} \cite{Sakai}. The main point that we wanted to make in the
present paper is that the complicated structure of the dressing factor
is in fact very natural and not in any way more complicated than for
the ordinary relativistic $O(4)$ model. The difference lies only in
the different expressions for the kernels appearing in the nested
Bethe ansatz - and this leads, almost accidentally, to more
complicated formulas. So in a way the fact that the
dressing kernel appears to be tightly linked to the nested structure
is not related to some `stringy' phenomenae specific for AdS/CFT but
is rather a generic fact present also in quite simple relativistic
integrable field theories. We must emphasize, however, that this
discussion does not preclude the existence of different formulations
of the AdS theories in line of \cite{MP,Kaz} related to various ways
of quantizing the string using e.g. lightcone or covariant methods.

Finally we would like to emphasize that there remains a lot to be
understood. Foremost is why the observation of the present paper seems
to work at all. Furthermore, in our construction there is a set of discrete
choices\footnote{Both for the AdS case and for relativistic theories.}
which gives 
the correct formula. At the present moment we do not understand what
physical/mathematical principle picks the correct choice. We hope
that the precise understanding of these issues will shed new light on
the structure of integrable field theories in general and on the
rather mysterious features of the AdS dressing factor.

\bigskip

\noindent{\bf Acknowledgments.} We thank Volodya Kazakov, Adam Rej,
Pedro Vieira and the group at IFT Madrid for interesting discussions.  
This work has been supported in part by Polish Ministry of Science and
Information Technologies grant 1P03B04029 (2005-2008), RTN network
ENRAGE MRTN-CT-2004-005616, and the Marie Curie ToK KraGeoMP (SPB
189/6.PRUE/2007/7). 

\appendix

\section{Fourier transform conventions}

There are a few different conventions for the definitions of the
convolution and the Fourier transform. To make the discussion more clear
and help the reader follow the computations we present our conventions
and some basic relations used in the whole paper.  
\begin{itemize}
\item Convolution
\eq
(f* g)(x)=\int_{-\infty}^{+\infty}f(x-y)g(y)\mbox{d}y
\eqx
\item Fourier transform
\eq
\hat f(k)=\int_{-\infty}^{\infty}f(x)e^{i k x}\mbox{d}x
\eqx
\item Inverse Fourier transform
\eq
\check f(x)=\frac{1}{2\pi}\int_{-\infty}^{\infty}f(k)e^{-i k x}\mbox{d}k
\eqx
\item Fourier transform of the convolution 
\eq
\widehat{(f*g)}(k)=\hat{f}(k)\hat{g}(k)
\eqx
\item Fourier transform of the inverse of the function
  ($(f*f^{-1})(x)=\delta(x)$) 
\eq
\widehat{f^{-1}}(k)=\frac{1}{\hat{f}(k)}
\eqx
\item Fourier transform of the kernel $\phi_a$ defined in
  (\ref{phi.ofour})
\eq
\widehat{\phi}_a(k)=-(\mbox{\rm sgn } a)\, e^{-|ak|}
\eqx
\end{itemize}

\section{Proof of (\ref{b})}

To find the formula for the dressing factor we have to solve equations
(\ref{eq.o2n})--(\ref{abethe5}). The Fourier transforms of these
equations take the following form  
	\eq
	\label{eqns}
	 \left\{ \begin{tabular}{ll}
	$\hat{\rho}_{1}+\hat{\rho}_{1}^{h}=\hat{\phi}_{-\frac{1}{2}}
           \hat{\rho}_{3}+\hat{\phi}_{1}\hat{\rho}_{1}$,&\\ 
	$\hat{\rho}_{2}+\hat{\rho}_{2}^{h}=\hat{\phi}_{-\frac{1}{2}}
           \hat{\rho}_{3}+\hat{\phi}_{1}\hat{\rho}_{2}$,&\\
	$\hat{\rho}_{3}+\hat{\rho}_{3}^{h}=\hat{\phi}_{-\frac{1}{2}}
           \hat{\rho}_{1}+\hat{\phi}_{-\frac{1}{2}}\hat{\rho}_{2}+
           \hat{\phi}_{-\frac{1}{2}}\hat{\rho}_{4}+\hat{\phi}_{1}
           \hat{\rho}_{3}$,&\\  
	$\hat{\rho}_{k}+\hat{\rho}_{k}^{h}=\hat{\phi}_{-\frac{1}{2}}
           \hat{\rho}_{k+1}+\hat{\phi}_{-\frac{1}{2}}\hat{\rho}_{k-1}+
           \hat{\phi}_{1}\hat{\rho}_{k}$,& for $3<k\leq n$\\
	\end{tabular}\right.
	\eqx
	
Our claim is that the solution of (\ref{eqns}) is of the form 
	\eq\label{v}
	\hat{\rho}_{k-1}=\hat{\phi}_{-\frac{1}{2}}
        \frac{1+(\hat{\phi}_{-1})^{k-3}}{
          1+
          (\hat{\phi}_{-1})^{k-2}}\hat{\rho}_{k}+f_{k}(\hat{\rho}^{h}_{1},
        \ldots,\hat{\rho}^{h}_{k-1}),\ \ \ \mbox{for }3\leq k \leq n
	\eqx
where $f_{k}$ for each $k=1,\ldots,n$ is a function which does not
depend on any of $\hat{\rho}_{l}$ for $l=1,\ldots, n+1$. We will prove
this formula by induction. 

\noindent A straightforward computation shows that our claim is
true for $k=3$. Let us assume that (\ref{v}) holds for a given $k$. Then
solving  
	\eq
	\left\{ \begin{tabular}{l}
	$\hat{\rho}_{k}+\hat{\rho}_{k}^{h}=\hat{\phi}_{-\frac{1}{2}}
          \hat{\rho}_{k+1}+\hat{\phi}_{-\frac{1}{2}}\hat{\rho}_{k-1}+
          \hat{\phi}_{1}\hat{\rho}_{k}$\\
	$\hat{\rho}_{k-1}=\hat{\phi}_{-\frac{1}{2}}
        \frac{1+(\hat{\phi}_{-1})^{k-3}}{
          1+
          (\hat{\phi}_{-1})^{k-2}}\hat{\rho}_{k}+f_{k}(\hat{\rho}^{h}_{1},
        \ldots,\hat{\rho}^{h}_{k-1})$
	\end{tabular}\right.
	\eqx
we get (\ref{v}) taking into account the simple equality
$\hat{\phi}_{-1}=(\hat{\phi}_{-\frac{1}{2}})^2$. Now
we can plug it into the equation for the dressing factor
(\ref{e.dro2n}):
\eq
\phi_\sg =-\f{1}{2}\, \phi_{\frac{1}{2}}\ast \Phi
\eqx
and get 
	\eq
	\hat{\phi}_{\sigma}=\f{1}{2}\hat{\phi}_{-1}
        \frac{1+(\hat{\phi}_{-1})^{n-2}}{
          1+(\hat{\phi}_{-1})^{n-1}},
        \ \ \ \mbox{for } n \geq 2 
	\eqx

\end{document}